\documentclass[journal,transmag]{IEEEtran}
\ifCLASSINFOpdf
   \usepackage[pdftex]{graphicx}
\else
   \usepackage[dvips]{graphicx}
\fi
\usepackage{amsmath}
\usepackage{amsfonts}
\usepackage{algorithmic}
\usepackage{algorithm}
\floatname{algorithm}{Procedure}

\usepackage{stfloats}
\usepackage{subfigure}

\usepackage{multirow}

\begin{document}
%
\title{A Data Driven Approach for Resting-state EEG signal Classificition of Schizophrenia with Control Participants Using Random Matrix Theory}


\author{\IEEEauthorblockN{Haichun Liu\IEEEauthorrefmark{12},
TianHong Zhang\IEEEauthorrefmark{34}，
Yumeng Ye\IEEEauthorrefmark{12},
Changchun Pan\IEEEauthorrefmark{12},\\
Genke Yang\IEEEauthorrefmark{12},
JiJun Wang\IEEEauthorrefmark{34},
Robert C. Qiu\IEEEauthorrefmark{56},~\IEEEmembership{Fellow,~IEEE}}

\IEEEauthorblockA{\IEEEauthorrefmark{1}Key Laboratory of System Control and Information Processing, Ministry of Education of China,}
\IEEEauthorblockA{Shanghai Jiaotong University, Shanghai 200240, China.}
\IEEEauthorblockA{\IEEEauthorrefmark{2}Department of Automation,
Shanghai Jiaotong University, Shanghai 200240, China.}
\IEEEauthorblockA{\IEEEauthorrefmark{3}Shanghai Mental Health Center, Shanghai Jiaotong University School of Medicine, }
\IEEEauthorblockA{Shanghai Key Laboratory of Psychotic Disorders, Shanghai 200030, China}
\IEEEauthorblockA{\IEEEauthorrefmark{4}Brain Science and Technology Research Center, Shanghai Jiao Tong University, Shanghai 200240, China.}
\IEEEauthorblockA{\IEEEauthorrefmark{5}Department of Electrical Engineering,Shanghai Jiaotong University, Shanghai 200240, China.}
\IEEEauthorblockA{\IEEEauthorrefmark{6}Department of Electrical and Computer Engineering, Tennessee Technological University, Cookeville, TN 38505, USA.}
}

%



\IEEEtitleabstractindextext{%
\begin{abstract}
Resting state electroencephalogram (EEG) abnormalities in clinically high-risk individuals (CHR), clinically stable first-episode patients with schizophrenia (FES), healthy controls (HC) suggest alterations in neural oscillatory activity. However, few studies directly compare these anomalies among each types. Therefore, this study investigated whether these electrophysiological characteristics differentiate clinical populations from one another, and from non-psychiatric controls. To address this question, resting EEG power and coherence were assessed in 40 clinically high-risk individuals (CHR), 40 first-episode patients with schizophrenia (FES), and 40 healthy controls (HC). These findings suggest that resting EEG can be a sensitive measure for differentiating between clinical disorders.This paper proposes a novel data-driven supervised learning method to obtain identification of the patient’s mental status in schizophrenia research. According to Marchenko-Pastur Law, the distribution of the eigenvalues of EEG data is divided into signal subspace and noise subspace. A test statistic named LES that embodies the characteristics of all eigenvalues is adopted. different classifier and different feature(LES test function) are selected for experiments, we have shown that using von Neumann Entropy as LES test function combine with SVM classifier could obtain the best average classification accuracy during three classification among HC, FES and CHR of Schizophrenia group with EEG signal. It is worth noting that the result of LES feature extraction with the highest classification accuracy is around 90\% in two classification(HC compare with FES) and around 70\% in three classification. Where the classification accuracy higher than 70\% could be used to assist clinical diagnosis.
\end{abstract}

}

\maketitle

\IEEEdisplaynontitleabstractindextext

%
\IEEEpeerreviewmaketitle

\section{Introduction}
%
%
%
%
\IEEEPARstart{T}{he} analysis of the EEG signal has a long history. Being used as a diagnostic tool for nearly 70 years, it still resists a strict and objective analysis and its interpretation remains to be mostly intuitive and heuristic. In particular, the cross channel relations of the signal and its reference to the sources and hence to certain morphological and/or functional brain structures are not understood sufficiently and are a matter of intense research.\cite{Febo2008High}

Many methods have been developed with the aim of helping to understand the meaning of the EEG signal. They range from a visual inspection of the record by an experienced physician to sophisticated estimations which attempt to describe the signal using phase space methods or to model its sources by a set of oscillating electric dipoles.\cite{Lan2015Resting}\cite{Fraschini2014An} However none of these methods can be regarded as fully satisfactory.

The reason is that the EEG signal is an electric activity of the brain measured by electrodes (channels), which placed on the scalp is a superposition of electric signals which are produced by a synchronous activity of numerous neurons.\cite{Qibin2009Slice} The spatial propagation of the electric signal in the complex brain tissue is far from being straightforward. Moreover, various groups of neurons participate in various and sometimes independent tasks. All this together makes the resulting signal complex and difficult to interpret. It has to be stressed that the morphological and functional properties of the brain are not always the same. On the contrary, they can be different for different individuals. This makes the EEG signal dependent on the measured subject. Even if different persons perform the same tasks and the measurement is done under identical conditions the resulting signals could differ.

EEG possesses many special properties that would make it useful in this context, such as a high time resolution which opens a window to see the dynamics of the brain. Previous studies have observed that EEG provides important information about differences between individuals, as pertains to the anatomical and functional traits of their brains. An individual person’s EEG, furthermore, is both stable and specific.\cite{Vijayalakshmi2014Change} That is to say, EEG provides small intra-personal differentiation and large inter-personal differentiation, which is why it is ide{}al for biometrics.

EEG-based biometric systems can be basically organized into two different states: task-related state and resting state. In this paper we are interested in using resting state EEG. There are two reasons for this. First, evidence indicates that electrical activities’ resting state organizes and coordinates neuronal functions . Second, certain tasks cannot be performed by certain group of people, e.g., Attention Deficit Disorder (ADD) or handicapped patients. Resting state EEG includes Resting-state EEG with open eyes (REO) and Resting-state EEG with closed eyes (REC). In this paper, we will consider closed eyes (REC) conditions.\cite{Fouad2015Brain}

The common procedure of EEG-based biometrics involves data collection, preprocessing feature extraction, and pattern recognition.\cite{Wang2014Emotional} However, resting state EEG lacks a task related feature, thus making it difficult to manually design the best feature to extract. In this paper, our aim is to analyze the EEG covariances using tools of the random matrix theory. To this aim it is necessary to define a matrix ensemble based on the EEG data. Here we use the fact that the brain is nonstationary. This means that the tasks it is performing change with the time. Hence the structure of the covariance matrix changes. A typical EEG measurement lasts about 5 mins. This time interval can me divided into roughly thousands not overlapping stationary windows over which the covariance matrix is evaluated. In such a way we get a set of thousands covariance matrices. These matrices will represent the ensemble we will work with.

It is known that random matrix theory describes correctly the spectral statistics of certain complex systems.\cite{Qiu2016Smart} It is useful, in particular, when one deals with systems which are chaotic but display at the same time certain wave (or coherent) properties. An EEG signal is a multivariate time series. Being measured on the scalp, the activity of a particular cerebral area influences the results of several EEG channels. Hence the activity of a given cortex area leads to correlations between the signal measured on different electrodes. The object to be analyzed here is the simplest one describing this covariance, namely, the covariance matrix.

The random matrix theory, based on the random variables as the elements of the matrix, through the comparison of random multidimensional time series statistical properties, can reflect the correlation of real data and the degree of deviation from a random distribution properties, and reveal the behaviour characteristic of the whole actual data.\cite{He2017A} It is a unique angle of view that is different from the traditional methods, RMT is widely applied in the fields of financial, physical, biological statistics, computer science, etc. Based on the RMT, in the study of large radar antenna array signal transmission, it’s found that eigenvalue distribution which uses antenna as sensor nodes to construct random matrix, can well describe the correlation between each antenna nodes flow, and can make rapid response to the emergence of abnormal signals in the transmission process.\cite{Qiu2013Cognitive} In the financial field, through the analysis, the researchers found the trend of America stock maximum eigenvalue of correlation matrix in 1962-1979 years, analysed American stock character when the market crashed in 1987.

This paper proposes a novel data-driven supervised learning method to obtain identification of the patient's mental status in schizophrenia research. First, random matrix theories (RMT) are briefly introduced as the solid mathematic foundations for RMMs. Built on RMMs, the major data processing ingredients LES designs are systematically studied. 1) LES are high dimensional statistical indicators; with different test functions, LES gain insight into the systems from different perspectives. Moreover, some theoretical values related to LESs are predictable as the reference points via the latest theorems. 2) Besides, combined entropy with LES and try to use entropy inequalities to illustrate the real data information. It is a deep research; we just give a brief mention. In general, based on RMT and the big data applying architecture, this paper presents a series of work associated with the theorems of LES, a briefly discussion about information entropy, the indicators derived from LES, and the visualization of the results. Case studies, with real EEG data, validate the proposed method, and related theories and theorems.

The paper is organized as follows: The first part consists of an introduction to EEG and Prodromal Phase of Schizophrenia. The second part gives a general overview of EEG data source and processing. The third part, ”Materials and Methods”, describes the subjects and the experiment, as well as the topology of the RMT. Finally, the results, discussion and conclusion are detailed in Sections 4 and 5.

\section{Data source and processing}

\begin{figure}[!hbt]
\centering
\includegraphics[height = 8cm]{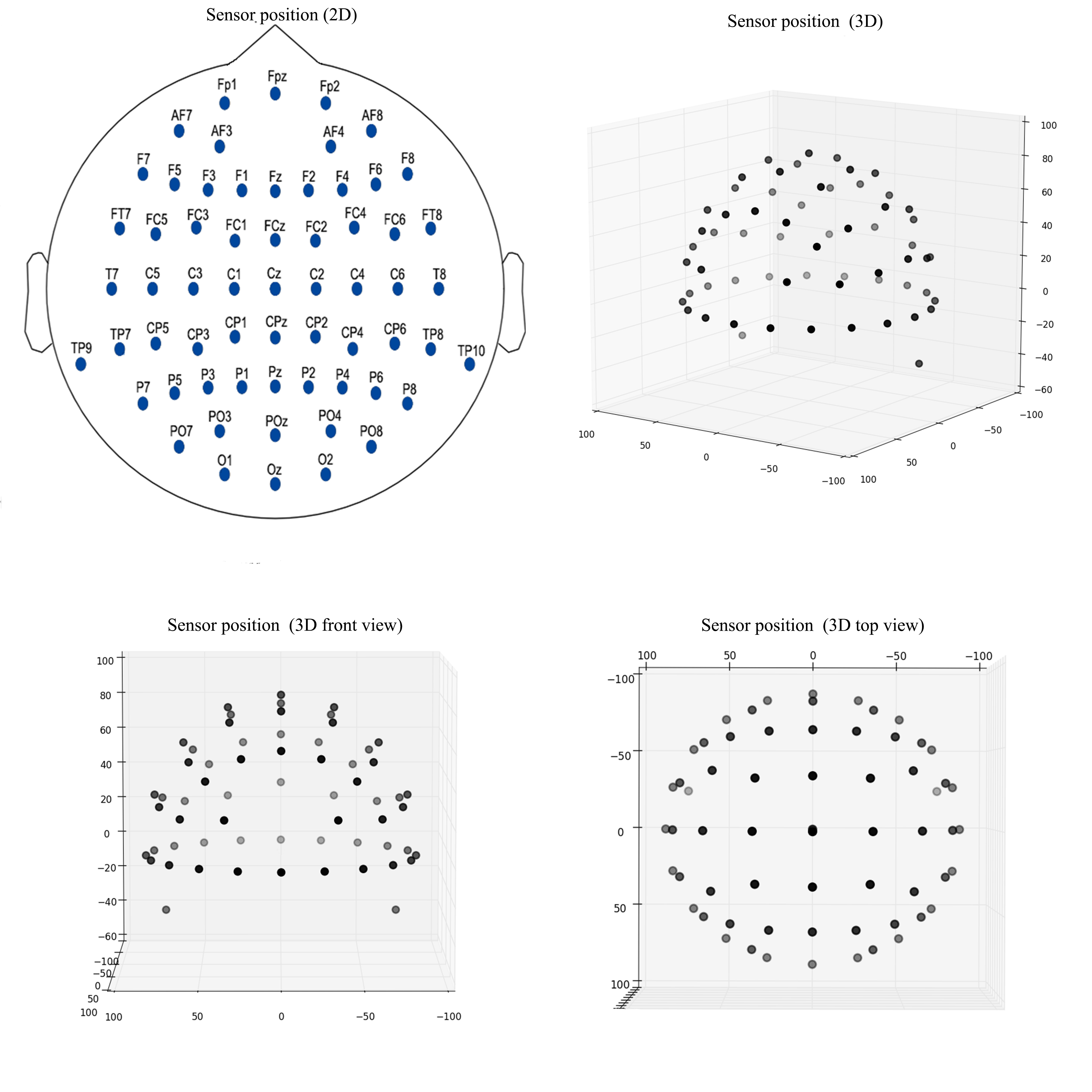}
\caption{EEG sensor}\label{fig: EEG sensor}  
\end{figure}

The data of this paper used was provided by one metal health center hospital of Shanghai. Continuous EEG activity was recorded while subjects were seated comfortably with their eyes closed. EEG recordings took place between 09.00 and 17.00. Participants were monitored for electroencephalographic signs of somnolence for the whole duration of the recording (5 mins). Recordings were conducted with 64 Ag/AgCl electrodes arranged according to a modified 10/20 system (additional positions: AF7, AF3, AF4, AF8, F5, F1, F2, F6; F10, FT9, FT7, FC3, FC4, FT8, FT10, C5, C1, C2, C6, TP7, CPz, TP8, P5, P1, P2, P6, PO3, POz, PO4), and referenced to FCz. Electrode impedance was kept below 5 kΩ. As EEG raw signal is contaminated with noise and artifacts of external and physio- logic origin, a 0.1– 50 Hz bandpass filter and ocular correction ICA were applied offline. The space position of the EEG sensor is shown in Fig. \ref{fig: EEG sensor}.

\begin{figure}[!hbt]
\centering
\includegraphics[height = 4cm]{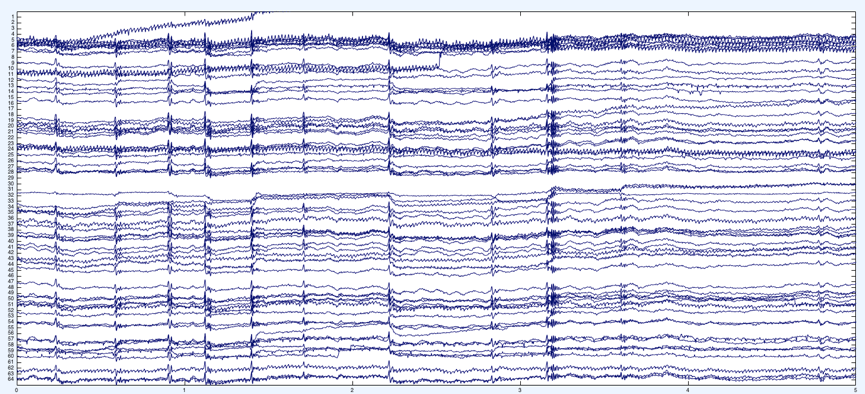}
\caption{EEG data}\label{fig: EEG data}  
\end{figure}

We describe this kind of time series data in the form of matrix, which dimension is N and sample size is T. N means numbers of the sensor which is 64 Ag/AgCl electrodes in this paper, and T mean period of the time series data which sample time is 5 mins and sampling frequency is 1000Hz. Participants were 10 HR, and 10 HC subjects, so that the data form of each subject is $64\times300000(N=64, T=300000)$, which is illustrated in Fig. \ref{fig: EEG data}

\section{Mathematics and Methods}

\subsection{Modeling of Large Dimensional Data As Random Matrix}
In multivariate statistics, we describe a random sample of $p$-dimensional random vectors $x_1,\ldots,x_T \in \mathbb{C}^N$, then we form the data matrix $X=(x_1,\ldots,x_T) \in \mathbb{C}^{N\times T}$, which naturally, is a random matrix need the analysis of high-dimensional data. The most remarkalbe fact is that $N$ and $T$ are large and comparale. The possibility of arbitrary sample size $N$ makes the classical statistical tools infeasible. We are asked to consider the asymptotic regime.
$$ N \to \infty , T \to \infty, but \frac{N}{T} \to c \in(0,\infty)  $$
The so-called random matrix theory is the natural answer to this question.\cite{Zhang2014Data}

\subsection{Random Matrix Theories}
Compared with traditional probability theory, random matrix is defined as a random variable for the elements of the matrix. If the random matrix dimension tends to infinity, it is called Large Random Matrix. Let rectangular matrix $X(N\times T)$, when $N\to\infty$,$T\to\infty$, and $N/T=c\to\sigma$, $\sigma$ is fixed. The large random matrix row and column tends to infinity, but the rank of the ratio is kept constant. Then, the large random matrix empirical spectral distribution (ESD) of this case has many proven characteristics, like Semicircular Law， Marchenko-Pastur (M-P) Law, and the latest Circle Law.

Random matrix theory according to the element distribution characteristics can be divided into: Gauss random matrices, Wishart random matrix, Haar-unitary random matrix. The research method is based on a variety of mathematical theory: the classical limit theory, free probability, mathematical transform, information theory. Several theorems which are given below are more important in RMT, this study is based on the following theorem.

Consider a sample covariance matrix of the form
$$ M=N^{-1}XX^{H} $$
where $X$ is a matrix whose entries $\left \{ X_{jk} \right \}_{j=1,\ldots,N;k=1,\ldots,T}$ are independent random variables, satisfying the conditions
$$E\left \{ X_{jk} \right \}=0,E\left \{ (X_{jk})^{2} \right \}=1$$
Let$\left \{ \lambda_{j} \right \}_{i=1}^{N}$be eigenvalues of M.

\subsubsection{Marchenko-Pastur Law}
Random matrix theory deals with statistical properties of the eigenvalues $\lambda_j$ of the covariance matrix M
$$ M|\lambda _j\left. \right \rangle=\lambda _j|\lambda _j\left. \right \rangle$$
where $|\lambda _j\left. \right \rangle$ being the corresponding eigenvectors. The simplest property of the eigenvalues family related to a matrix ensemble is the eigenvalue density $\rho (\lambda )$.It counts the mean number of eigenvalues contained in an interval $(a,b)$.
$$ \left \{ \lambda _j;\lambda _j \in (a,b) \right \}=\int_{a}^{b}\rho (\lambda )d\lambda $$
According to the above distribution entries of matrix $X$, we get an ensemble of random matrix which was mathematically studied by Marchenko and Pastur. In particular, the density $\rho(\lambda)$ of the ensemble is known adn given by the formula, if $N/T=c<1$, whose density is given by:
\begin{equation}
\begin{aligned}
\rho _c(\lambda )=\frac{1}{2\pi c \lambda }\sqrt{(b-\lambda )(\lambda -a)}\\
where, a=(1-\sqrt{c})^2,b=(1+\sqrt{c})^2
\end{aligned}
\end{equation}
where N being the number of channels, T is the number of the sampling points.This Marchenko-Pastur Law is the analogue of Wigner's Semicircle Law in this setting of multiplicative rather than addtive symmetrization, the assumption fo Gaussian entries may be significantly relaxed.

\begin{figure*}[htbp]
\centering
\subfigure[Distribution of the eigenvalues of Simulation data] {\includegraphics[height=2.5in]{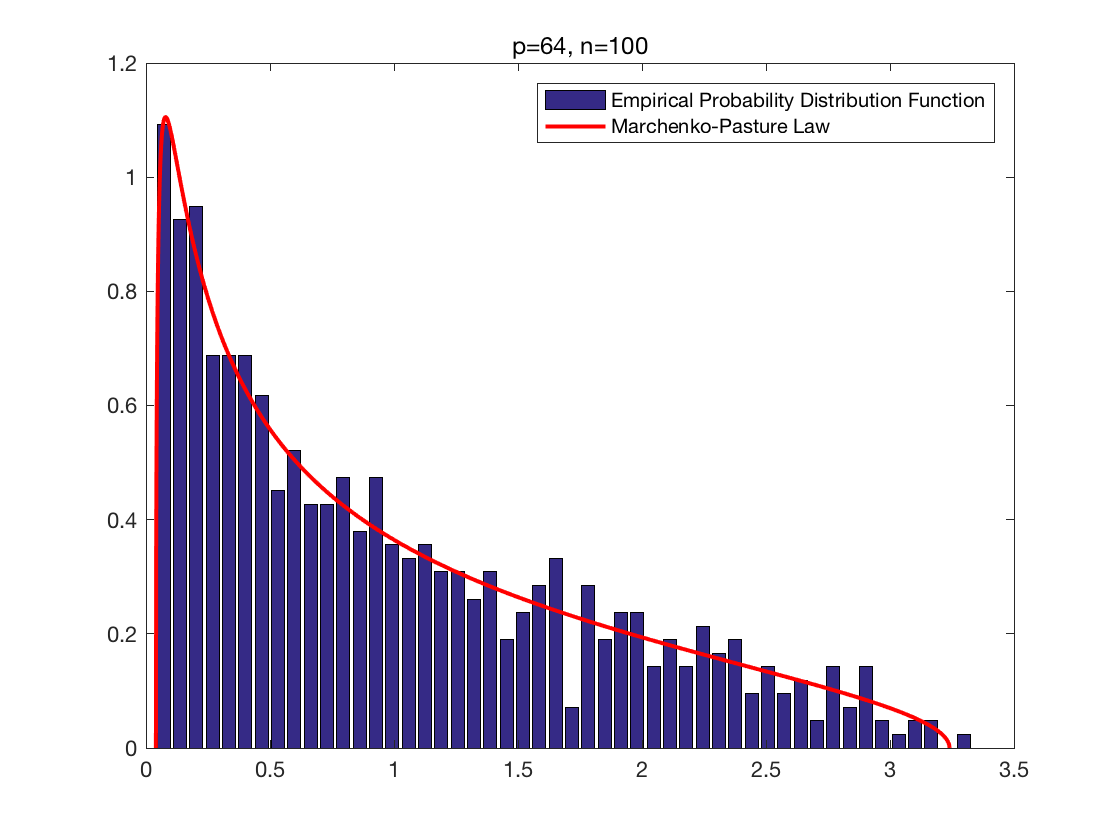}}
\subfigure[Distribution of the eigenvalues of EEG data] {\includegraphics[height=2.5in]{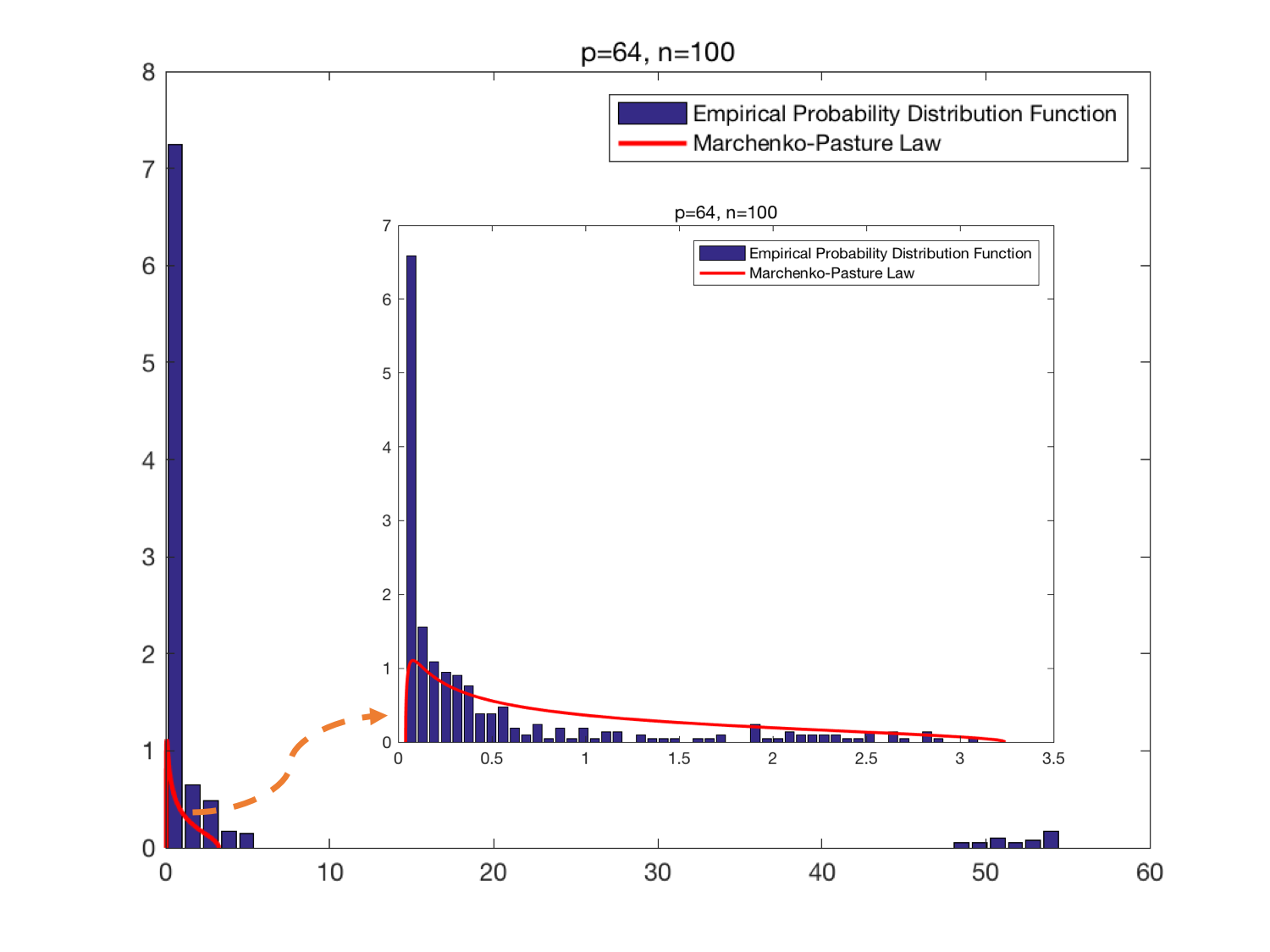}}
\caption{ Empirical Probability Distribution and Limit Probability Distribution of Eigenvalues of Data Covariance Matrix }
\label{fig: mplaw}
\end{figure*}

With the definition of this spectral density, it needs the independence of the data. In consideration of EEG is not a random signal, on the contrary, it is synchronous, which is related to the corresponding brain activity. We can expect that the spectral density does not follow the prediction of Marchenko-Pastur Law, and it will be related to the human activities. But fortunately, real EEG data is not completely inconsistent with this spectral distribution. The result is plotted in Fig. \ref{fig: mplaw}. To keep the result simple to comprehend we randomly choose the data frame from one object, because these three group of data shows an approximate result.

The blue bar in Fig. \ref{fig: mplaw} (a) is the empirical spectral distribution of the eigenvalues of $N^{-1}XX^{H}$ where $X$ is an $N\times T, N=64, T=100$ random Gaussian matrix, the red curve is the Marchenko-Pastur law with density function. There are some differences between real and simulated data. In Fig. \ref{fig: mplaw} (b), on the head of the distribution, there is a certain number of eigenvalues near zero, and on the tail of the distribution, there are several large eigenvalues which are outliers. When we magnify the eigenvalues empirical spectral distribution(blue bar) below the limit spectral distribution(MP Law, red curve), we are surprised to find that the distribution of tempirial and limit of this part coincides with a certain extent. According to the definition of the random matrix, the part below M-P Law shows that the data conforms to the characteristics of random noise, which is named "noise subsapce"，but the noise subspace here is not fully fit the theoretical value, because this part also contains a few part of the information. Other parts including the eigenvalues near zero and outliers show information of the data, which is named "signal subspace". Eigenvalues near zero means correlation of data, and outliers can be affected by actual activities. To derive profound and subject-independent information from EEG data, a test statistic that embodies the characteristics of all eigenvalues should be adopted.

\subsubsection{Linear Eigenvalue Statistics}
Definition:  Since the pioneer work of Wigner it is known that if we consider the linear eigenvalue statistic corresponding to any continuous test function $\varphi$\cite {O2016Central}
$$N_{n}[\varphi ]=\sum_{j=1}^{n}\varphi (\lambda _{j})$$
The law of Large Numbers is an important step in studies of eigenvalue distributions for a certain random matrix ensemble. Then $n^{-1}N_{n}[\varphi]$converges in probability to the limit:\cite{Shcherbina2011Central}
$$\lim_{n\to\infty}n^{-1}N_{n}[\varphi]=\int\varphi (\lambda )\rho (\lambda )d\lambda $$
where $\rho(\lambda)$ is the probability density function (PDF) of the eigenvalues. 

Consider a sample covariance matrix with entries of $X$, Let the real values test function $\varphi$ satisfy condition $\left \| \varphi  \right \|_{3/2+\varepsilon }<\infty(\varepsilon >0)$. Then $N_n^0[\varphi]$ in the limit $N,T\to\infty, N/T\to c \geq 1$ converges in distribution to the Gaussian random variable with zero mean and the variance, which is the CTL for $M$:
\begin{equation}
\begin{aligned}
V[\varphi ]&=\frac{1}{2\pi ^2}\int_{a_+}^{a_-}\int_{a_+}^{a_-}\\
&(\frac{\Delta \varphi }{\Delta \lambda })^2\frac{(4c-(\lambda _1-a_m)(\lambda_2-a_m))d\lambda_1d\lambda_2}{\sqrt{4c-(\lambda_1-a_m)^2}\sqrt{4c-(\lambda_2-a_m)^2}}\\
&+\frac{k_4}{4c\pi ^2}(\int_{a_+}^{a_-}\varphi (\mu )\frac{\mu -a_m}{\sqrt{4c-(\mu -a_m)^2}}d\mu )^2
\end{aligned}
\end{equation}
where $\frac{\Delta \varphi }{\Delta \lambda}=\frac{\varphi(\lambda_1)-\varphi(\lambda_2)}{\lambda_1-\lambda_2},k_4=X_4-3$ is the fourth cumulant of entries of  $X, a_\pm =(1\pm\sqrt{c})^2, a_m=\frac{1}{2}(a_++a_-)$

Similarly, we can design other test functions to obtain diverse LES as the indicators, and some theoretical values. Here we list some classical test functions, show in Table \ref{table: several popular test function}.
\begin{table*}[bp]
\caption{several popular test function}
\begin{center}
\begin{tabular}{lccc}
\hline
Soure&Original Form&Equivalent LES&Test Function\\
\hline
Folklore-LRT&$Tr(A)-\log \det(A)-Tr(I_N)$&$\sum_{i=1}^{n}(\lambda_i-\log \lambda_i-1)$&$x-\log x-1$\\
Wasserstein Distance&$Tr(A-2\sqrt{A}+I_N)$&$\sum_{i=1}^{n}(\lambda_i-2\sqrt{\lambda_i}+1)$&$x-2\sqrt{x}+1$\\
Nagao(1973)&$Tr(A-I_N)^2$&$\sum_{i=1}^{n}(\lambda_i-1)^2$&$(x-1)^2$\\
von Neumann Entropy&$-Tr(A\log \det A)$&$\sum_{i=1}^{n}(-\lambda_i\log\lambda_i)$&$-x\log x$\\
\hline
\end{tabular}
\end{center}
\label{table: several popular test function}
\end{table*}%

As mentioned in the previous subsection, LES is a test statistic containing the information of the eigenvalues. We employ a statistical test methods, like analysis of variance (ANOVA), which are applied on the selected features to examine the pre-defined hypothesis. Results can be reported using p values to denote the significance of difference in signatures between/among factors for testing hypothesis.
\begin{figure}[!hbt]
\centering
\includegraphics[height = 6cm]{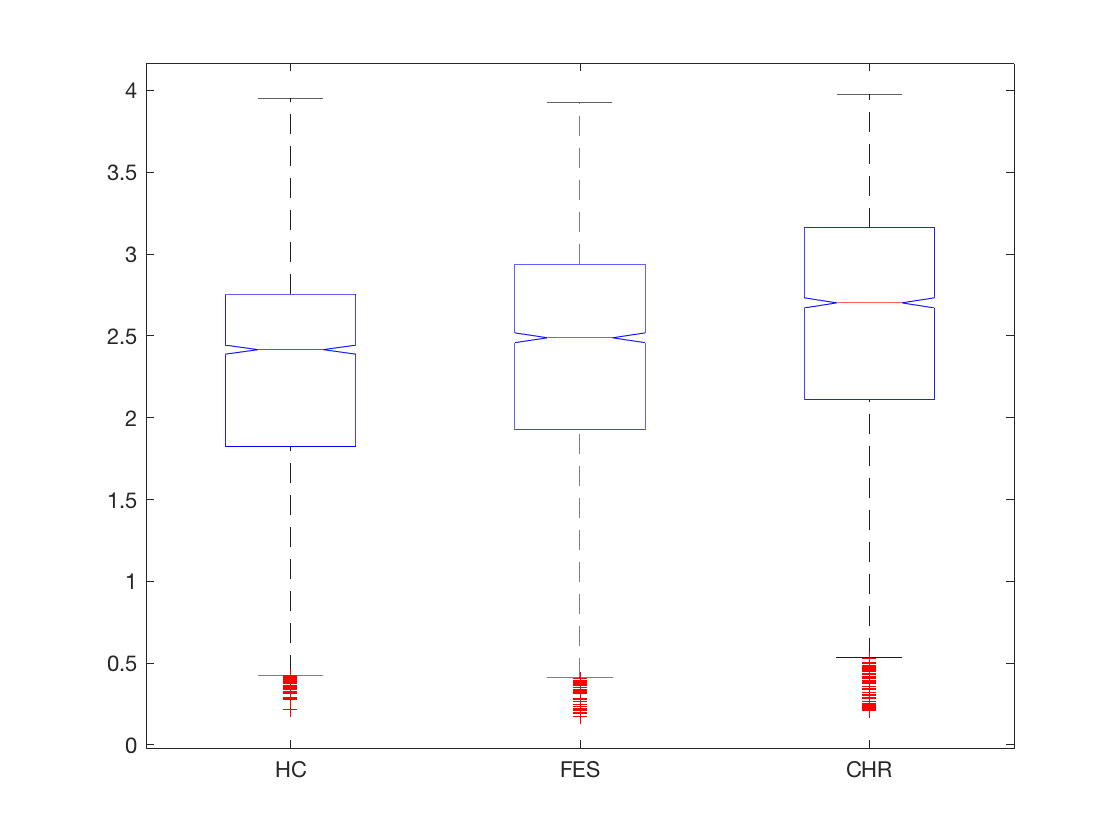}
\caption{Analysis of variance of three groups of people}\label{fig: anova1}  
\end{figure}

\begin{table}[htbp]
\caption{p-value of three groups of people}
\begin{center}
\begin{tabular}{ccccc}
\hline
 &ALL&HC\&FES&FES\&CHR&HC\&CHR\\
\hline
p-value&$1.26\times 10^{-5}$&$2.67\times 10^{-7}$&$1.48\times 10^{-12}$&$5.21\times 10^{-3}$\\
\hline
\end{tabular}
\end{center}
\label{table: pvalue}
\end{table}%

We pick the form of von Neumann Entropy as a test LES statistic, p-value table and data box plot is shown in Table \ref{table: pvalue} and Fig. \ref{fig: anova1}. The p-value of three states of Schizophrenia and each pairs of three states is small enough (p \textless 0.05), p-value that is smaller than the significance level indicates that at least one of the sample means is significantly different from the others. From the result of analysis of variance，there should be an appropriate form of LES test function could be the feature of classification among different kinds of states of Schizophrenia.

\subsubsection{Information Entropy}
Shannon defined the entropy $H$ of a discrete random variable $X$ with possible values $\{x_1,\ldots,x_n\}$ and probability mass function $P(X)$ as:
\begin{equation}
H(X)=E[I(X)]=E[-\ln(P(X))]
\end{equation}
Here E is the expected value operator, and I is the information content of $X$.  $I(X)$is itself a random variable.

The information entropy can explicitly be written as:
\begin{equation}
H(X)=\sum_{i=1}^{n}P(x_i)I(x_i)=-\sum_{i=1}^{n}P(x_i)\log_bP(x_i)
\end{equation}
where b is the base of the logarithm used. Common values of b are 2, Euler number e, and 10, and the unit of entropy is Shannon for b =2, while the units of entropy are also commonly referred to as bits.\cite{Nadal2011Statistical}

As mentioned before we can design other test functions to obtain diverse LES as the indicators, but it is real hard to find an appropriate test function. We might have noticed that after the standardization of the matrix, the sum of eigenvalues of matrix is equal to 1, and each eigenvalues is greater than 0, less than 1, which is just a kind of probability.
\begin{equation}
\sum_{i=1}^{n}\lambda_i=1,0<\lambda_i<1(1<i<n)
\end{equation}
As a result, it is natural to think of the formula of information entropy could be the test function of LES. In consideration of LES is with no physically meaning, but information entropy is kind of description of the uncertainty of the source, which could be linked to the actual.

\subsubsection{Universality Principle}
Akin to CLT, universality refers to the phenomenon that the asymptotic distributions of various covariance matrices (such of eigenvalues and eigenvectors) are identical to those of Gaussian covariance matrices. These results let us calculate the exact asymptotic distributions of various test statistics without restrictive distributional assumptions of matrix entries. The presence of the universality property suggests that high- dimensional phenomenon is robust to the precise details of the model ingredients. For example, one can perform various hypothesis tests under the assumption that the matrix entries not Gaussian distributed but use the same test statistic as in the Gaussian case.
The data of real systems can be viewed as a spatial and temporal sampling of the random graph. Randomness is introduced by the uncertainty of spatial locations and the system uncertainty. Under real life applications, we cannot expect the matrix entries follow i.i.d. distribution. Numerous studies based on both simulations and experiments, however, demonstrate that the Ring Law, M-P Law and LES are universally followed. In such cases, universality properties provide a crucial tool to reduce the proofs of general results to those in a tractable special case: the $i.i.d.$ case in our paper.

\subsection{Methods}
\begin{figure}[!hbt]
\centering
\includegraphics[width = 8cm]{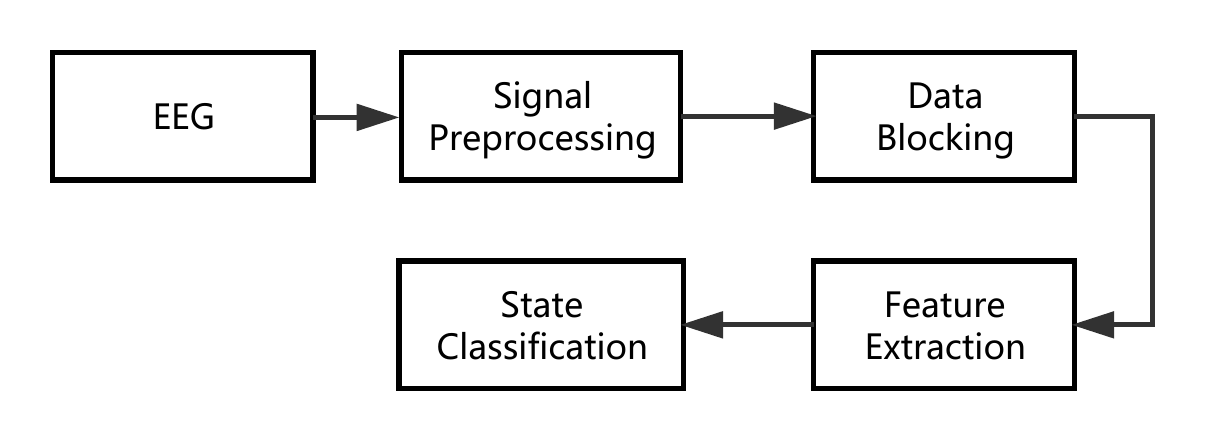}
\caption{EEG procedures}\label{fig: EEG procedures}  
\end{figure}

The EEG data we got from the experiments mentioned before were analyzed through several procedures, including signal preprocessing, data blocking, feature extraction and state classification, as shown in Fig. \ref{fig: EEG procedures}, where the data blocking and feature extraction are the most important parts in this paper.

\subsubsection{Signal Preprocessing}
The basic approach to signal analysis is to get proper information from the signal by applying the best suitable process. The process method used in this study is Fourier transform.

Spectral analysis of a signal involves decomposition of the signal into its frequency components. In other words, the original signal could be separated into its sub-spectral components by using spectral analysis methods. Among spectral analysis techniques, Fourier transform is considered to be the best transformation between time and frequency domains because of it being time shift invariant. The Fourier transform pairs are expressed as:
\begin{equation}
X(k)=\sum_{n=0}^{N-1}x(n)W_N^{kn}
\end{equation}
\begin{equation}
x(n)=\frac{1}{N}\sum_{n=0}^{N-1}X(k)W_N^{-kn}
\end{equation}
where $W_N=e^{-j(2\pi/N)}$ and $N=length[x(n)]$.

\subsubsection{Data Blocking}
\begin{figure}[!hbt]
\centering
\includegraphics[width = 8cm]{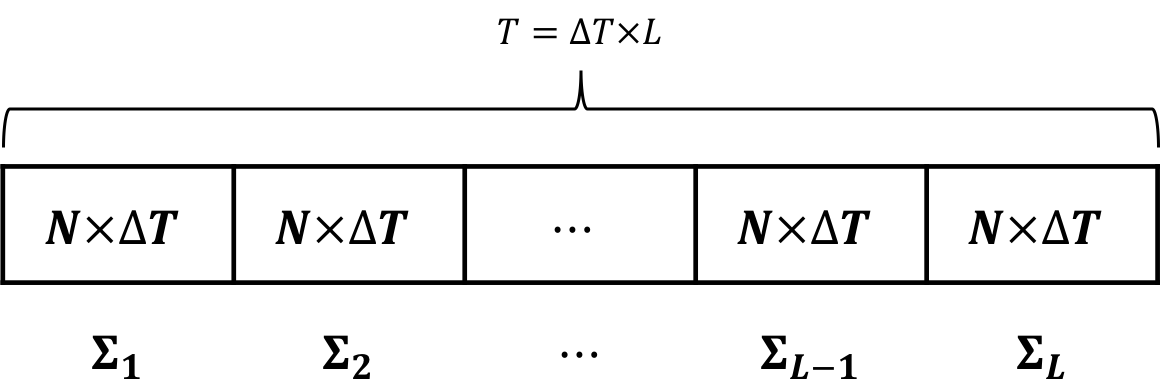}
\caption{random matirx flow}\label{fig: random matirx flow}  
\end{figure}

We try to turn big EEG data into tiny data for practical use, some lately advanced data driven estimators could be adopted to obtain state evaluation without knowledge of the medical parameters or connectivity.\cite{Chu2016Massive} In consideration of the characteristic of data is an extremely long time series, Fig. \ref{fig: random matirx flow} illustrates the conceptual representation of the structure of the massive streaming EEG data. More specifically, the original data form of each subject is a $N\times T$ matrix. Let $\Sigma_i=N\times \Delta T$ and L be the window size and sampling number($T=\Delta T \times L$), respectively. Then the time series data was divided into a sequence of random matrix:
\begin{equation*}
\left \{ \Sigma_1,\Sigma_2,\ldots,\Sigma_L \right\}
\end{equation*}

\subsubsection{Feature Extraction}
The main task of feature extraction is to derive the salient features which can map EEG data into different states. Feature extraction is related to dimensionality reduction, when the input data to an algorithm is too large to be processed and it is suspected to be redundant. It is fundamental to study the spectral analysis of large dimensional random matrix concerns the central limit theory for LES. The main reason is because many important statistics in multivariate statistics analysis can be expressed as functionals of the empirical spectral distribution of some random matrix. Because we are concerned with the mapping of eigenvalue, sample covariance matrix and LES(kind of dimensionality reduction) were parts of feature extraction.

{\bfseries Sample Covariance Matrix} which deals with the question of how to approximate the actual covariance matrix on the basis of sample from the multivariate distribution.The sample covariance matrix is an unbiased and efficient estimator of the covariance matrix if the space of covariance matrix is viewed as extrinsic convex cone in $R^{p\times p}$.

{\bfseries Linear Eigenvalue Statistics} in this study represent the similar function to PCA, which involves reducing the amount of resources required to describe a large set of EEG data in statistics.

\subsubsection{State Classification}
The computed features are then fed to the classifier for classification between different states of the human brain. We have used the following classifiers in our study.

{\bfseries SVM classifier:} Support vector machines commonly known as SVMs are considered the best state of the art classifiers having lower complexity as compared to other classifiers like neural networks and fuzzy classifiers. SVM based upon as concept to find the hyper plane which can able to classify the data to the separate classed with possibility of maximum margin. A linear SVM can also classify nonlinear problems by using kernel trick with little amount of complexity increase. 

{\bfseries Decision Tree:} It is a top-down recursive construction method. Each internal node represents a test on an attribute, each edge represents a test result, the leaf node represents a kind of class or class distribution, the top node is the root node. DT is a typical classification method, which is to approximate the value of discrete function.

{\bfseries Random Forest:} It is a random way to build a forest, there are a lot of decision tree in the random forest, and no covariance between each decision tree. Random forest could be used for classification and regression.

{\bfseries Naive Bayes:} It is based upon Bayes theory and it makes an assumption that each attribute of given class is independent of the values of other attributes. Class conditional independence is based upon this assumption.

\renewcommand{\algorithmicrequire}{\textbf{Input:}} 
\renewcommand{\algorithmicensure}{\textbf{Output:}}
\begin{algorithm}
\caption{EEG data classification using random matrix theory}
\begin{algorithmic}[1]
	\REQUIRE Raw EEG data matrix X($N\times T$) , which is a time series data matrix of $N$ dimension
	\STATE Turning big EEG data into tiny data, select appropriate $\Delta T$ to turning raw data into a sequence of $L$ sample matrices $\mathcal{X}$($N\times \Delta T$), $X=\mathcal{X} \times L$
	\FOR{Each sample matrix $\mathcal{X}$}
		\STATE {Standardization: whose entries satisfying the conditions $E\left \{ X_{jk}^{(n)} \right \}=0,E\left \{ (X_{jk}^{(n)})^{2} \right \}=1$}
		\STATE {Obtaining a sample covariance matrix of matrix $\mathcal{X}$ of the form $M=N^{-1}\mathcal{X}\mathcal{X}^{H}$}
		\STATE {Select appropriate test function $\varphi$ to calculate LES as the feature $\zeta_i,1\leq i\leq L $}
	\ENDFOR
	\STATE {The obtained features $\left \{ \zeta _1,\zeta _2,\ldots,\zeta _L \right \}$ are used as input to various machine learning algorithms}
	\ENSURE Classification results for each states
	\end{algorithmic}
\label{procedure:1}
\end{algorithm}

\section{Cases and Discussion}
In oder to perform a more reliable classification process, we constructed a training set and test set from raw EEG data, the cross validation method is presented in this study. Training set consists of 80\% samples which randomly selected from raw data, the remaining part for the test set, first used the training set to train the classifier, and then use the test set to validate the trained model, finally repeat this process several times to get statistical information of observations. On this account, as a performance index evaluation of classifier. As mentioned above, several classification method were adopted in this study includes SVM, KNN, Naive Bayes, Decision Tree and Random Forest.

\subsection{Comparison of various extracted features}
The extraction of feature is extremely important since it seeks to reduce the dimensionality of the data that are synthesized in the form of feature vector, preserving and highlighting the useful information from the raw data. The quality of the feature selection directly affects the success rate of classification, and the computational effort required, the balance between success and efficiency should be taken into account.

\subsubsection{Effect of statistical features}
Statistical features are responsible for summarizing the windows in a global view of data\cite{Borges2017Feature}, which can be analyzed as follows:

\begin{table}[htp]
\caption{Summarized Description of Statistical Features}
\begin{center}
\begin{tabular}{ll}
\hline 
Name&Mathematical Expression\\
\hline
Harmonic mean&$\frac{N}{\sum_{j=1}^{N}\frac{1}{d_j}}$\\
Standard deviation&$\sqrt{\frac{\sum_{j=1}^{N}(d_j-mean)^2}{N-1}}$\\
Mean deviation&$\frac{\sum_{j=1}^{N}\left | d_{ij}-mean_i \right |}{N}$\\
Kurtosis&$\frac{\frac{1}{N}\sum_{j=1}^{N}(d_j-mean)^4}{\left [ \frac{1}{N}\sum_{j=1}^{N}(d_j-mean)^2 \right ]^2}$\\
Root mean square&$\sqrt{\frac{1}{N}\sum_{j=1}^{N}\left | d_j \right |}$\\
Peak value&$max(d_j)$\\
Difference between max and min&$max(d_j)-min(d_j)$\\
\hline
\end{tabular}
\end{center}
\label{default}
\end{table}%

The characteristics of each electrode can be obtained through these statistical features during the sampling period. There are 64 electrodes in all, each electrode could obtain seven statistical features, then we could get $64\times 7$ features. These features are trained through variety of machine learning methods:

\begin{table}[htp]
\caption{Classification accuracies of statistical features using various classifier}
\begin{center}
\begin{tabular}{lcc}
\hline
Classifier&HC\&FES&HC\&FES\&CHR\\
\hline
SVM&$79.16\pm 6.21\%$&$62.83\pm 8.54\%$\\
KNN&$73.25\pm 7.33\%$&$57.33\pm 9.71\%$\\
Navie Bayes&$75.78\pm 6.47\%$&$57.97\pm 8.09\%$\\
Decision Tree&$76.16\pm 6.01\%$&$59.11\pm 7.83\%$\\
Random Forest&$78.25\pm 5.76\%$&$61.74\pm 7.12\%$\\
\hline
\end{tabular}
\end{center}
\label{table: classification without les}
\end{table}%

Table \ref{table: classification without les} shows the classification performance obtained by various classifier under serval statistical features in distinguishing two or three kind of states of Schizophrenia. It can be observed that the two-category classification problem between HC and FES has good classification effect, the accuracies of each classifier is more than 70\%, particularly the SVM method is closed to 80\%.  It should be point out one important aspect here that SVM classifier has three kinds of kernels, namely linear, polynomial and RBF, we choose RBF kernel to get the result in this study. We have already make a sample comparative experiment among these kernels, which definitely proves the robustness of the RBF over the polynomial and linear for these data sets, so that SVM with RBF kernel was chosen as the basic kernel in the remaining paper.

But when the third state(CHR) is put in the classifier, the accuracies turn to be worse, the success rate is only around 60\%. This means that the feature extraction method can not distinguish the three states well. The result less than 70\% has no reference value for clinical applications.

\subsubsection{Effect of LES features}
In this section, we roughly represent EEG data in terms of high-dimensional data vectors, geometrically, there are points in a high-dimensional vector space. Here random variables are taken into account simultaneously both spatially and temporally, the central idea is to treat the vectors as a whole, rather than multiple samples of random vectors. According to the procedure \ref{procedure:1}, we try to turning raw EEG data into many tiny data blocks, each block calculates a LES feature on the basis of some test function. For a random matrix $A\in \mathbb{C}^{n\times n}$  whose eigenvalues $\lambda _i, i=1,\ldots ,n$, are known as strong correlated random variables. We use concept in section 2 to show that several popular test functions in Table \ref{table: several popular test function}.

Different classifier and different feature(LES) are selected for experiments, and the classification results are contrasted and analyzed. For the result shown in Table \ref{table: classification with les}, It is worth noting that the result of LES feature extraction is better than traditional statistical feature extraction about 10\%, the highest classification accuracy is around  90\% in two classification and around 70\% in three classification which is very encouraging. It can be seen that the performance of classifiers used were satisfactory in distinguish between HC and FES, and using von Neumann Entropy as LES test function combine with SVM classifier could obtain the average classification accuracy of 73.31\% during three classification among HC, FES and CHR. Comparing with the result in Table \ref{table: classification without les}, It is obvious that two classification result between HC and FES could be relatively easy to get good results, where the classification success rate higher than 70\% could be used to assist clinical diagnosis. Different from statistical feature, based on LES feature

\begin{table}[h]
\caption{Classification accuracies of LES features using various classifier}
\begin{center}
\begin{tabular}{lccc}
\hline
Classifier&Test Function&HC\&FES&HC\&FES\&CHR\\
\hline
\multirow{4}{*}{SVM}&LRT&$88.34\pm 2.33\%$&$68.56\pm 2.98\%$\\
&W-Distance&$88.86\pm 2.78\%$&$69.85\pm 3.34\%$\\
&Nagao&$86.27\pm 3.27\%$&$67.73\pm 4.31\%$\\
&vN-Entropy&$91.16\pm 1.61\%$&$73.31\pm 2.23\%$\\
\hline
\multirow{4}{*}{KNN}&LRT&$76.89\pm 3.21\%$&$61.24\pm 4.87\%$\\
&W-Distance&$77.97\pm 3.44\%$&$61.36\pm 4.54\%$\\
&Nagao&$76.12\pm 4.05\%$&$59.54\pm 5.77\%$\\
&vN-Entropy&$78.33\pm 3.56\%$&$61.74\pm 4.06\%$\\
\hline
\multirow{4}{*}{Navie Bayes}&LRT&$75.82\pm 3.73\%$&$56.88\pm 5.43\%$\\
&W-Distance&$76.32\pm 3.64\%$&$57.91\pm 5.12\%$\\
&Nagao&$75.11\pm 4.39\%$&$56.93\pm 6.31\%$\\
&vN-Entropy&$76.59\pm 3.11\%$&$58.45\pm 4.67\%$\\
\hline
\multirow{4}{*}{Descision Tree}&LRT&$81.97\pm 2.98\%$&$65.23\pm 4.32\%$\\
&W-Distance&$82.41\pm 3.25\%$&$65.93\pm 3.94\%$\\
&Nagao&$81.16\pm 3.47\%$&$64.81\pm 4.77\%$\\
&vN-Entropy&$81.22\pm 2.93\%$&$65.76\pm 3.91\%$\\
\hline
\multirow{4}{*}{Radndom Forest}&LRT&$86.35\pm 3.32\%$&$67.45\pm 3.96\%$\\
&W-Distance&$89.63\pm 2.12\%$&$67.77\pm 3.59\%$\\
&Nagao&$86.31\pm 2.86\%$&$66.83\pm 3.65\%$\\
&vN-Entropy&$85.94\pm 2.66\%$&$68.39\pm 3.11\%$\\
\hline
\end{tabular}
\end{center}
\label{table: classification with les}
\end{table}%

If your training set is small, high bias/low variance classifiers (e.g., Naive Bayes) have an advantage over low bias/high variance classifiers (e.g., kNN ), since the latter will overfit. But low bias/high variance classifiers start to win out as your training set grows (they have lower asymptotic error), since high bias classifiers aren't powerful enough to provide accurate models. Though, that better data often beats better algorithms, and designing good features goes a long way. And if you have a huge dataset, your choice of classification algorithm might not really matter so much in terms of classification performance. The data sets used in this study are large, but not large enough to ignore the differences between various classification methods, in other words, between the various classification methods are not significantly different, but the differences cannot be ignored.

According to Table \ref{table: classification with les}, under the same feature selection, the results of SVM were better than those of other classifications. It is pretty obvious that the advantages of SVM is High accuracy, nice theoretical guarantees regarding overfitting, and with an appropriate kernel they can work well even if you're data isn't linearly separable in the base feature space, which is especially popular in classification problems where high-dimensional spaces are the norm. 

For a $p\times p$ random matrix $A$ with eigenvalues $\lambda _i, i=1,\ldots ,n$ linear spectral statistics of type:
\begin{equation*}
\frac{1}{p}\sum_{i=1}^{p}f(\lambda _i)={\rm Tr}\;f(A)
\end{equation*}
for various test function $f$ are of central importance in the theory of random matrix.Several popular statistical indicators were used to be test function in this study, each test function has its physical meaning, and the significance of selecting the test function is that it can give the physical explanation to LES. LES has statistical significance, and the test function has a physical meaning, such a combination, you can try to find the engineering interpretation of the classification results with the novel algorithm. The von Neumann Entropy performance the best results compare with other functions, There will be further discussion about physical meaning of entropy later in this study.

\subsection{Effect of different window size in LES feature}
Before, we used a fixed sampling period as the sampling base to do the identification, and obtained good classification results. 
In this section, we would like to find the other frequency length of EEG signal which is still able to preserve the identification information, and we will compare their performance differences. 
We divided each EEG signal sampling point into various portions($\Delta T$ from 100 to 5000) and then performed in the mentioned procedure.
Based on the results of the above comparative experiments, von Neumann Entropy is used as test function and SVM is selected for classification.

The classification results are shown in Table \ref{table: effect of different windows}.
From the table we can observe that as the length of sampling size became longer, 
different $\Delta T$ will affect the estimation accuracy of the sample covariance matrix, the maximum classification accuracy is located at $\Delta T = 200$.
In classical statistics, sample size is usually 3-8 times the dimension, this means the suitably range of $\Delta T$ should be from 150 to 500.
The results may suggest that the sampling size over which subjects can be individualized is around at 200.

\begin{table}[htp]
\caption{Effect of different windows in LES feature}
\begin{center}
\begin{tabular}{lcc}
\hline
$\Delta T$&HC\&FES&HC\&FES\&CHR\\
\hline
100&$87.71\pm 3.66\%$&$68.14\pm 4.74\%$\\
200&$91.16\pm 1.61\%$&$73.31\pm 2.23\%$\\
500&$88.96\pm 2.91\%$&$70.09\pm 3.13\%$\\
1000&$87.11\pm 2.63\%$&$68.65\pm 3.88\%$\\
2000&$85.32\pm 3.37\%$&$66.48\pm 3.75\%$\\
5000&$78.97\pm 3.94\%$&$60.19\pm 4.67\%$\\
\hline
\end{tabular}
\end{center}
\label{table: effect of different windows}
\end{table}%

\subsection{von Neumann entropy}
The von Neumann entropy is a generalization of the classical entropy (Shannon entropy) to the field of quantum mechanics. 
The von Neumann entropy is one of the cornerstones of quantum information theory. 
It plays an essential role in the expressions for the best achievable rates of virtually every coding theorem. 

For any quantum state described by a Hermitian positive semi-definite matrix $\rho$, the von Neumann entropy of $\rho$ is defined as
\begin{equation}
	S(\rho)=-{\rm Tr}(\rho \log \rho)
\end{equation}

For notation, here we prefer using bold upper-case symbols $X, Y, Z$ to represent random matrices. 
If $X$ is a symmetric (or Hermitian) matrix of $n \times n$ and $f$ is a bounded measurable function,
$f(X)$ is defined as the matrix with the same eigenvectors as $X$ but with eigenvalues that are the images by $f$ of those of $X$;
namely, if $e$ is an eigenvector of $X$ with eigenvalues $\lambda$, $Xe = \lambda e$, then we have $f(x)e=f(\lambda)e$.
For the spectral decomposition $X = UDU^H$ with orthogonal (unitary) and $D=diag (\lambda _1),\ldots , \lambda _n$ diagonal real, one has
\begin{equation}
	f(X)=Uf(D)U^H
\end{equation}
with $(f(D))_{ii}=f(\lambda _i),i=1,\ldots,n$. We could obtain
\begin{equation}
	S(X)=-{\rm Tr}(X\log X)=-\sum_{i=1}^{n}\lambda _i(X)\log \lambda _i(X)
\end{equation}
where $\lambda _i (X), i = 1,\ldots, n$ are eigenvalues of $X$.

The von Neumann entropy is a standard measure of entanglement between subsystems: 
which reaches its minimum 0 when the system is unentangled,and its maximum $\log n$ when the system is maximally entangled.

\section{Conclusion}
In conclusion, different classifier and different feature(LES test function) are selected for experiments, we have shown that using von Neumann Entropy as LES test function combine with SVM classifier could obtain the best average classification accuracy during three classification among HC, FES and CHR of Schizophrenia group with EEG signal. The classification performance would be improved as the size of training and data database becomes larger, in the future, the proposed biometrics system should be tested on a larger group and more classes of subjects, providing further identification of accuracy, robustness and applicability of the system. 

There is extensive research space for EEG data analysis of human psychiatric disorders. Our current work is focused on EEG spectral analysis. The prospective work on EEG data analysis could be the further data analysis based on Spatio-Temporal representation information of EEG signal. We can introduce a more suitable model to solve this problem.


%

%

\section*{Acknowledgment}

We are appreciated for department of EEG source imaging leaded by professor Jijun Wang and professor Chunbo Li from SHJC for data providing and discussion. We also grateful to Dr. Tianhong Zhang from SHJC for his expert collaboration on data analysis.

\ifCLASSOPTIONcaptionsoff
  \newpage
\fi



%


\bibliographystyle{IEEEtran}
\bibliography{EEG_reference}

\begin{thebibliography}{10}
\providecommand{\url}[1]{#1}
\csname url@samestyle\endcsname
\providecommand{\newblock}{\relax}
\providecommand{\bibinfo}[2]{#2}
\providecommand{\BIBentrySTDinterwordspacing}{\spaceskip=0pt\relax}
\providecommand{\BIBentryALTinterwordstretchfactor}{4}
\providecommand{\BIBentryALTinterwordspacing}{\spaceskip=\fontdimen2\font plus
\BIBentryALTinterwordstretchfactor\fontdimen3\font minus
  \fontdimen4\font\relax}
\providecommand{\BIBforeignlanguage}[2]{{%
\expandafter\ifx\csname l@#1\endcsname\relax
\typeout{** WARNING: IEEEtran.bst: No hyphenation pattern has been}%
\typeout{** loaded for the language `#1'. Using the pattern for}%
\typeout{** the default language instead.}%
\else
\language=\csname l@#1\endcsname
\fi
#2}}
\providecommand{\BIBdecl}{\relax}
\BIBdecl

\bibitem{Febo2008High}
F.~Cincotti, D.~Mattia, F.~Aloise, S.~Bufalari, L.~Astolfi, F.~D.~V. Fallani,
  A.~Tocci, L.~Bianchi, M.~G. Marciani, and S.~Gao, ``High-resolution eeg
  techniques for brain--computer interface applications,'' \emph{Journal of
  Neuroscience Methods}, vol. 167, no.~1, pp. 31--42, 2008.

\bibitem{Lan2015Resting}
M.~Lan, J.~W. Minett, T.~Blu, and W.~S.~Y. Wang, ``Resting state eeg-based
  biometrics for individual identification using convolutional neural
  networks,'' \emph{Conf Proc IEEE Eng Med Biol Soc}, vol. 2015, pp.
  2848--2851, 2015.

\bibitem{Fraschini2014An}
M.~Fraschini, A.~Hillebrand, M.~Demuru, L.~Didaci, and G.~L. Marcialis, ``An
  eeg-based biometric system using eigenvector centrality in resting state
  brain networks,'' \emph{IEEE Signal Processing Letters}, vol.~22, no.~6, pp.
  666--670, 2014.

\bibitem{Qibin2009Slice}
Q.~Zhao, C.~F. Caiafa, A.~Cichocki, L.~Zhang, and A.~H. Phan, ``Slice oriented
  tensor decomposition of eeg data for feature extraction in space, frequency
  and time domains,'' vol. 5863, pp. 221--228, 2009.

\bibitem{Vijayalakshmi2014Change}
R.~Vijayalakshmi, N.~Dasari, D.~Nandagopal, R.~Subhiksha, B.~Cocks, N.~Dahal,
  and M.~Thilaga, ``Change detection and visualization of functional brain
  networks using eeg data,'' in \emph{International Conference on Computational
  Science}, 2014, pp. 672--682.

\bibitem{Fouad2015Brain}
M.~M. Fouad, K.~M. Amin, N.~Elbendary, and A.~E. Hassanien, ``Brain computer
  interface: A review,'' \emph{Intelligent Systems Reference Library}, vol.~74,
  pp. 3--30, 2015.

\bibitem{Wang2014Emotional}
X.~W. Wang, D.~Nie, and B.~L. Lu, ``Emotional state classification from eeg
  data using machine learning approach,'' \emph{Neurocomputing}, vol. 129,
  no.~4, pp. 94--106, 2014.

\bibitem{Qiu2016Smart}
R.~C. Qiu and P.~Antonik, ``Smart grid and big data,'' 2016.

\bibitem{He2017A}
X.~He, Q.~Ai, R.~C. Qiu, W.~Huang, L.~Piao, and H.~Liu, ``A big data
  architecture design for smart grids based on random matrix theory,''
  \emph{IEEE Transactions on Smart Grid}, vol.~8, no.~2, pp. 674--686, 2017.

\bibitem{Qiu2013Cognitive}
R.~Qiu and M.~Wicks, \emph{Cognitive Networked Sensing and Big Data}.\hskip 1em
  plus 0.5em minus 0.4em\relax Springer Publishing Company, Incorporated, 2013.

\bibitem{Zhang2014Data}
C.~Zhang and R.~C. Qiu, ``Data modeling with large random matrices in a
  cognitive radio network testbed: Initial experimental demonstrations with 70
  nodes,'' \emph{Eprint Arxiv}, 2014.

\bibitem{O2016Central}
S.~O'Rourke and D.~Renfrew, ``Central limit theorem for linear eigenvalue
  statistics of elliptic random matrices,'' \emph{Journal of Theoretical
  Probability}, vol.~29, no.~3, pp. 1121--1191, 2016.

\bibitem{Shcherbina2011Central}
M.~Shcherbina, ``Central limit theorem for linear eigenvalue statistics of the
  wigner and sample covariance random matrices,'' \emph{Journal of Mathematical
  Physics Analysis Geometry}, vol.~14, no.~69, pp. p{\'a}gs. 64--108, 2011.

\bibitem{Nadal2011Statistical}
C.~Nadal, S.~N. Majumdar, and M.~Vergassola, ``Statistical distribution of
  quantum entanglement for a random bipartite state,'' \emph{Journal of
  Statistical Physics}, vol. 142, no.~2, pp. 403--438, 2011.

\bibitem{Chu2016Massive}
L.~Chu, R.~C. Qiu, X.~He, Z.~Ling, and Y.~Liu, ``Massive streaming pmu data
  modeling and analytics in smart grid state evaluation based on multiple
  high-dimensional covariance tests,'' \emph{IEEE Transactions on Big Data},
  vol.~PP, no.~99, pp. 1--1, 2016.

\bibitem{Borges2017Feature}
F.~A.~S. Borges, R.~A.~S. Fernandes, I.~N. Silva, and C.~B.~S. Silva, ``Feature
  extraction and power quality disturbances classification using smart meters
  signals,'' \emph{IEEE Transactions on Industrial Informatics}, vol.~12,
  no.~2, pp. 824--833, 2017.

\end{thebibliography}

%











\end{document}